# Nanoscale ice fracture by molecular dynamics simulations

Amir Afshar, Jing Zhong, David S. Thompson[2] and Dong Meng[1]
*Mississippi State University, Starkville, MS, 39762, USA*

**In this work, we conducted molecular dynamics simulations to study the fracture mechanism of ice crystals in a bulk phase and at ice-ice interfaces at the atomistic scale. We show that there exists a narrow disordered interfacial layer between two *Ih* ice structures. The width of the interfacial layer is determined to be about the size of two water molecules. Upon deformation, the stress response of ice at interface show significantly anisotropic behaviors depending on the direction of deformation. Bulk-like behavior is observed when direction of deformation being orthogonal to the direction of interfacial plane. Significantly smaller fracture stress and yield strain occurs if the deformation is along interfacial plane. This result illustrates the dominant role played by the small amount of disordered water molecules at interface in altering mechanical strength of an interfacial structure.**

## I. Nomenclature

| | | |
|---|---|---|
| $\varepsilon$ | = | Lennard-Jones potential well |
| $\sigma$ | = | Lennard-Jones diameter of atoms |
| $r_{ij}$ | = | distance between the atoms |
| $e$ | = | proton charge |
| $\epsilon_0$ | = | permittivity of vacuum |
| q | = | molecular charge |
| $d_{OM}$ | = | distance of dummy particle from oxygen atom |
| k | = | Boltzmann constant |

## II. I. Introduction

Ice accretion is an important concern in aeronautical and aerospace industries. It occurs when supercooled liquid water droplets in the atmosphere are brought into contact with solid surfaces (such as aircraft bodies). The change in conditions of heat conductance, fluid dynamics, and interfacial energy breaks the metastable equilibrium state of liquid droplets turning them into solid ice, and subsequently leads to a rapid ice build-up on surfaces[1]. The consequences of ice accretion are often disastrous, leading to unfavorable aerodynamics performance and in several cases fatal accidents. Extensive research efforts have been devoted to design strategies for anti-icing and de-icing. One of the important aspects is to identify relevant parameters that are responsible for ice adhesion strength[2-3], which is manifested through fracture process of ice at ice-substrate interfaces. Studying fracture mechanism of ice-substrate interfaces at the nanoscale plays a particularly important role in further our understanding on this complicate phenomenon. It has been believed that the non-bonded interactions, including van der Waals forces and electrostatic interactions both have dominant effects on ice adhesion mechanism[4-5]. It is also demonstrated that a thin (~nm) layer of disordered molecules at interfaces can remarkably reduce ice adhesion strength[6-7]. However, access to the nanoscale mechanics through experimental measurements is limited by their temporal and spatial resolutions. It is the main objective of this study to use molecular dynamics (MD) simulations to shed lights on the nanoscale fracture mechanism of ice at interfaces, by offering atomistic details on the ice-substrate interfacial structure, bonding strength, and strain-stress distributions during fracture initialization and propagation. The findings of this study bring us a step closer to identifying the relevant parameters that controls ice-substrate adhesion strength. The developed simulation protocols allow further systematic investigations on ice fracture at various types of interfaces.

## III. Simulation Model and Method

The TIP4p/ice water model [7] is used in our simulations considering that it properly reproduces ice and water properties at low temperatures, and particularly the melting temperature of solid-liquid phase transition. TIP4p/ice model exhibits a bulk melting point of 272.2K that is close to the experimental value of 273.13K, while as a comparison TIP4p model has a melting temperature of 232K[19]. The water model consists of two parts, the Lennard-Jones potential and the Columbic potential as shown in Equation (1) and (2), respectively. Values for the parameters used in the two equations are listed in Table 1.

---

1. Address: The Swalm School of Chemical Engineering, Mississippi State University; email: dm2596@msstate.edu
2. Address: Department of Aerospace Engineering, Mississippi State University; email: dst@ae.msstate.edu

$$u_{ij}^{LJ} = 4\varepsilon \left[ \left(\frac{\sigma}{r_{i,j}}\right)^{12} - \left(\frac{\sigma}{r_{i,j}}\right)^6 \right]$$

(1)

$$u_{ij}^e = \frac{e^2}{4\pi\epsilon_0} \sum_i \sum_{j \neq i} \frac{q_i q_j}{|\mathbf{r}_{ij}|}$$

(2)

**Table 1. Parameters of the TIP4P water model[21]**

| Model | $\frac{\varepsilon}{k}(K)$ | $\sigma(\text{Å})$ | $q_H$ (e) | $d_{OM}$ (Å) |
|---|---|---|---|---|
| TIP4p/Ice | 106.1 | 3.1668 | 0.5897 | 0.1577 |

The Large-scale Atomic/Molecular Massively Parallel Simulator (LAMMPS)[9], a classical Molecular Dynamics (MD) simulation package, is used to perform all the simulations in this study. All simulations are conducted under the NPT ensemble -- temperature and pressure of the system are kept constant at T=250K and P=1.0bar using the Nosé–Hoover thermo- and barostat with coupling time constants of 0.4ps and 4.0ps, respectively. The MD time-step for integrating equation of motions was set to be 2.0fs. The Lenard-Jones interaction, and short-range part of the Coulombic potential are both truncated at 9.5 Angstrom (about three times the size of one water molecule). The long-range part of Coulombic interactions are evaluated using the particle-particle particle-mesh (PPPM) method[10]. During simulations, molecular structure of water molecules is kept rigid using the SHAKE algorithm. Periodic boundary conditions are applied in all three dimensions of the simulation box.

Fracture mechanism of two systems are investigated in this work. In the "reference" system uniform *Ih* (the proton disordered phase) ice crystal structure spans the simulation box of dimension $Lx$ = 4.7nm, $Ly$ = 5.1nm, and $Lz$ = 10.8nm and a total 8064 water molecules. In the "interface" system, two pieces of *Ih* ice crystals are brought into contact through the basal-prism interface along the x dimension as shown in Figure 2(b). The simulation box has a dimension of $Lx$=14.2nm, $Ly$=10.3, $Lz$=5.5nm, and contains a total number of 24640 water molecules. Before fracture simulation, both systems are equilibrated at T=250K and P=1bar. For the "reference" system, equilibration is concluded as the Oxygen-Oxygen radial distribution function calculated from our simulations reproduces equilibrated values reported in literatures [7] (Figure 1(a)). For the "interface" system, equilibration is monitored by tracking the number of *solid*-like water molecules (see Appendix for the method of determining the state of a water molecule) as function of simulation time, which develops a plateau when equilibrium is reached as shown in Figure 2(a). Starting from equilibrated structures (Figure 1(b)-(c) and 2(b)), fracture simulations are conducted by applying a constant strain rate of 5% per nanosecond along the direction of interest while keeping dimensions of other two directions constant, until fracture of ice is completed. The corresponding stress tensor component $P_{ii}$ is reported as the "stress" in the stress-strain plot as shown in Figure 3.

## IV. Results and Discussion

In this section, we outline the preliminary findings from the ice fracture simulations in the bulk phase (the "reference" system) and at an ice-ice interface (the "interface" system). Comprehensive analysis will be presented in the 2018 AIAA conference at Atlanta. Figure 3 shows the stress-strain responses upon deformation. For the "reference" system, deformations along *x* and *y* directions (the blue and green curves in Figure 3) correspond to deformations along the "prism" and "basal" plane of a *Ih* ice crystal structure, respectively. The corresponding stress-strain curves are mostly identical except near the yielding point, with pulling along the "prism" plane giving a higher yield stress at slightly longer strain ratio. Figure 4 (a) and (b) shows the structural changes during fracture initialization and propagation. It appears that fracture of a uniform *Ih* ice crystal initializes by first generating regions with local structural disordering of a length scale of tens of water molecules. Strain localization (fracture) occurs afterwards within these regions, and then propagate by cleaving hydrogen bonds. To further understand effects of different types of interactions on the overall stress, stress components from non-electrostatic and electrostatic interactions are plotted in Figure 3 (b)-(d) for the case of deformation along the *x* direction (results of other cases are qualitatively similar). It shows that the stress response is mainly due to non-electrostatic interactions, as the electrostatic component stays relatively constant before and after fracture. Interestingly, the stress component due to long-range electrostatic interactions Figure 3 (d) changes little in magnitude during and after fracture, except near the yield point. Further analysis of this observation may offer insight from a different aspect into the mechanism of fracture initialization inside ice crystals.

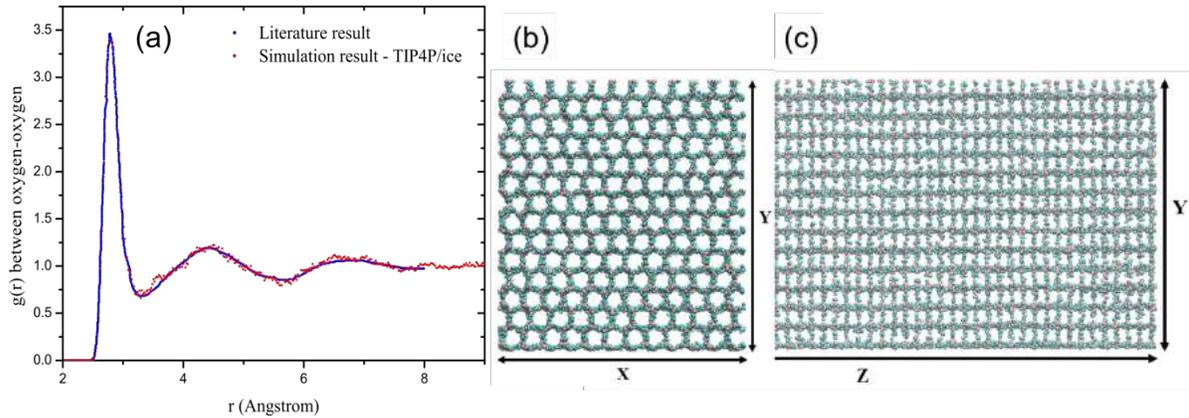

**Fig. 1** Equilibration of the "reference" system: (a) comparison of the Oxygen-Oxygen radial distribution function from our simulations and data from literature; (b)-(c) simulation snapshots taken at t=30ns of the equilibration stage, showing the plane views of the equilibrated uniform *Ih* ice structure.

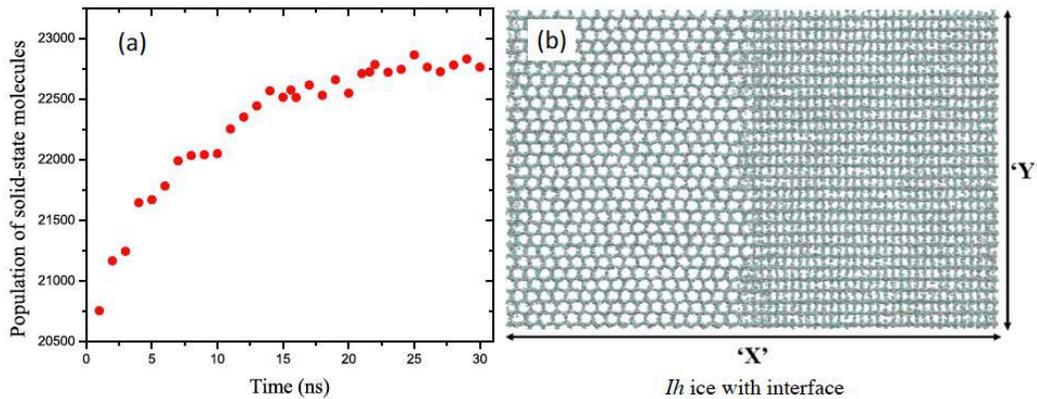

**Fig. 2** Equilibration of the "interace" system: (a) number of solid-like water molecules as function of simulation time; (b) a simulation snapshot taken at t=30ns of the equilibration stage, showing the plane view of the equilibrated *Ih* ice structure with an interface along *x* direction.

Deformations of the "interface" system are conducted along the *x* and *y* directions as indicated in Figure 2(b). In the case of deforming along the *x* direction, the two pieces of *Ih* ice structures experience stretching along the "basal" and "prism" plane, respectively. If deformation is uniformly affine, a stress-strain response compositing a "basal" and "prism" behavior (the blue and green curves in Figure 3) would be expected. However, Figure 3 shows that the corresponding stress-strain response (the red curve) is significantly weaker in terms of both modulus and yield stress. This result suggests that non-affine deformation occurs at an early stage, which implies that the narrow interfacial region shown in Figure 2 (b) tends to localize strain deformation, and eventually leads to fracture initialization and propagation. This conjecture is further proved by the snapshots in Figure 4 (c) showing the structural changes during facture. Interestingly, in the case of deformation along *y* direction, fracture apparently does *not* initialize at interface but *within* one of the *Ih* ice structure as shown in Figure 4(d). As the result, the corresponding stress-strain response resembles that of pure *Ih* ice fracture along the "basal" plane.

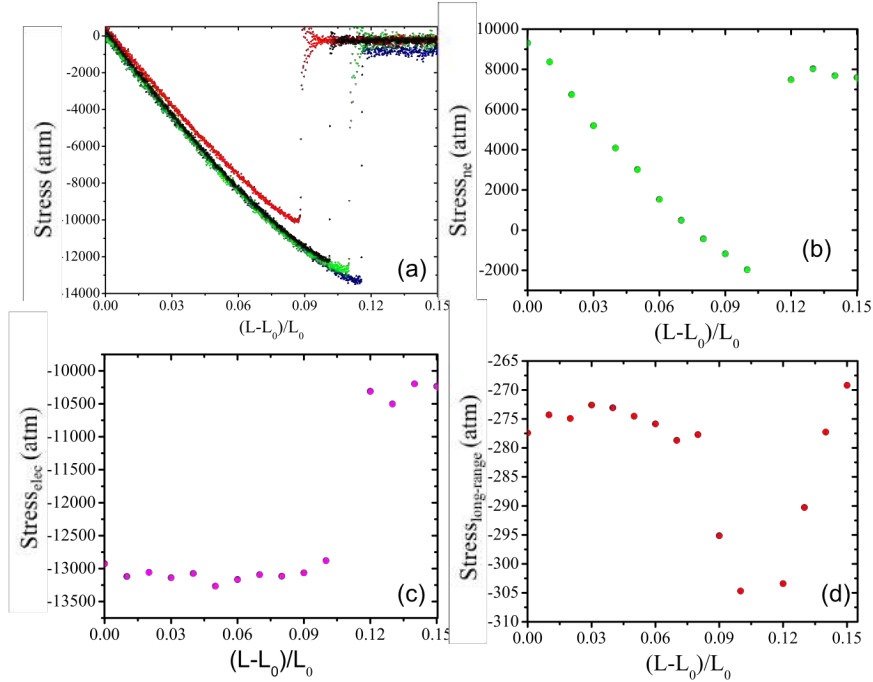

**Fig. 3** (a) Stress-strain response upon deformation along *x* (blue) and *z* (green) direction of the "reference" system, and along *x* (red) and *y* (black) direction of the "interface" systems. Response of (b) the non-electrostatic component, (c) the electrostatic component, and (d) the long-range electrostatic component of the total stress, upon deformation along *x* direction of the "reference" system.

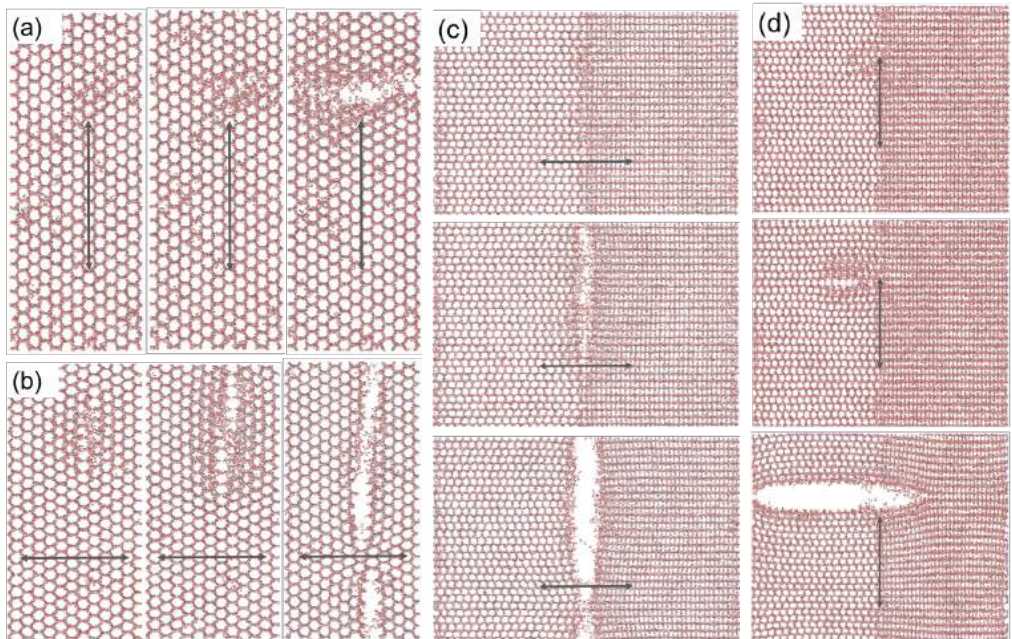

**Fig. 4** Snapshots from fracture simulations near the stress yielding point, showing structural evolutoins during fracture initialization and propagation in (a)-(b) the "reference" system, and (c)-(d) the "interface" system. Direction of deformation are inidicated by the arrows in each plot.

## Acknowledgments

The authors gratefully acknowledge the National Aeronautics and Space Administration (NASA) for the sponsor and financial support in the project of "Multiscale Physics-Based Modeling of Impact Ice Adhesion" (Grant No. NNX16AN20A).

## V. Appendix

In distinguishing the *solid*- and *liquid*-like particles, the local bond order parameter of each water molecule is calculated according to Equation (3) and (4)[27], and the threshold of $q_6 > 0.37$ is used as the criterion for determining a molecule to be *solid*-like.

$$q_l(i) = \sqrt{\frac{4\pi}{2l+1} \sum_{m=-l}^{l} \left(q_{lm}(i)\right)^2}, \text{ where } l = 6 \tag{3}$$

$$q_{lm}(i) = \frac{1}{N_b(i)} \sum_{j=1}^{N_b(i)} Y_{lm}(r_{ij}) \tag{4}$$